# Generation and detection of dissipationless spin current in MgO/Si bilayer

Paul C Lou[1], and Sandeep Kumar[1,2*]

[1] Department of Mechanical Engineering, University of California, Riverside, CA

[2] Materials Science and Engineering Program, University of California, Riverside, CA

**Abstract**

Spintronics is an analogue to electronics where spin of the electron rather than its charge is functionally controlled for devices. The generation and detection of spin current without ferromagnetic or exotic/scarce materials are two the biggest challenges for spintronics devices. In this study, we report a solution to the two problems of spin current generation and detection in Si. Using non-local measurement, we experimentally demonstrate the generation of helical dissipationless spin current using spin-Hall effect. Contrary to the theoretical prediction, we observe the spin-Hall effect in both n-doped and p-doped Si. The helical spin current is attributed to the site-inversion asymmetry of the diamond cubic lattice of Si and structure inversion asymmetry in MgO/Si bilayer. The spin to charge conversion in Si is insignificant due to weak spin-orbit coupling. For the efficient detection of spin current, we report spin to charge conversion at the MgO (1nm)/Si (2 μm) (p-doped and n-doped) thin film interface due to Rashba spin-orbit coupling. We detected the spin current at a distance of >100 μm, which is an order of magnitude larger than the longest spin diffusion length measured using spin injection techniques. The existence of spin current in Si is verified from coercivity reduction in Co/Pd multilayer due to spin-orbit torque generated by spin current from Si.

Spintronics devices require generation, transport, detection and manipulation of pure spin current. Si is earth-abundant and is considered to be an ideal material for semiconductor spintronics. Spin injection in Si has been experimentally demonstrated by tunneling from a ferromagnetic electrode across a thin insulator[1-4] with spin diffusion length of up to ~ 6 μm[5]. The long spin diffusion length at room temperature makes it an ideal spin channel (transport) material. The inverse spin-Hall effect (ISHE) has been demonstrated in p-Si[6] although the spin-Hall angle is extremely small. The spin-Hall effect (SHE)[7] [8] [9] is considered to be an efficient method for generation of pure spin currents using an electric field. The intrinsic SHE has been proposed to exist in some p-type semiconductors including GaAs, Ge and Si. This spin current is proposed to be quantum in nature, hence dissipationless[10]. SHE was observed in gallium arsenide (GaAs) using optical detection techniques, Kerr microscopy and a two-dimensional light-emitting diode [11, 12]. Si is an indirect band-gap semiconductor, so optical methods are not applicable for studying the SHE in Si. In addition, the spin-orbit coupling in Si is very small (44meV), and intrinsic ISHE may not produce a measurable signal. The experimental evidence of the SHE has been reported in p-Si using magneto-thermal transport measurements, but the mechanism of SHE is not clearly demonstrated. The long spin diffusion length and SHE satisfy the two requirements of spintronics devices: spin transport and spin current generation. In the absence of a reliable spin detection mechanism, the Si spintronics may not be practically realizable. In addition, the scientific understanding of the mechanism of the SHE is essential for manipulation of spin current. In this work, we present the experimental proof of dissipationless spin current in Si while efficient spin to charge conversion is achieved at the MgO/p-Si interface. We utilize the non-local measurement technique similar to the experimental setup proposed by Abanin et al.[13, 14]. We hypothesized that the SHE in Si will lead to spin accumulation as shown in Figure 1 a, and the non-local measurement may allow us to characterize the mechanism of spin current in Si.

(Figure 1)

In this study, our goal is to uncover the dissipationless spin current and the spin diffusion length in Si can be up to 6 µm[5, 15]. If dissipationless spin current exists then the specimen dimensions have to be significantly larger than it. We designed the experimental setup accordingly. For the experimental measurements, we developed a freestanding Hall bar MEMS structure using 2 µm p-Si (0.001-0.005 Ω cm) as shown in Figure 1 b with channel width of 20 µm. To fabricate the experimental devices, we take a commercially available silicon on insulator (SOI) wafer with 0.001-0.005 Ω cm with device layer of 2 µm. Using photolithography and deep reactive ion etching (DRIE), we pattern the front side (device layer) with specimen and electrodes. Then, we etch the back side of the wafer underneath the sample area to have the freestanding specimen using DRIE. The MgO thin film deposition is carried out using RF sputtering. The specimen is made freestanding to avoid any vertical temperature gradient, which may lead to electric potential due to anomalous Nernst effect (ANE)/Nernst effect in case of magneto-transport measurements. We carried out the measurement inside a Quantum Design physical property measurement system (PPMS). We undertake the temperature-dependent non-local resistance measurement for a current of 2 mA (37 Hz) applied across J3 and measured non-local resistances of a p-Si specimen as a function of temperature are shown in Figure 1 c. The measured non-local resistances are in agreement with the van der Pauw theorem[16] ($R_{NL} = R_{sq} e^{\frac{-\pi L}{w}}$ where $R_{sq} = \frac{\rho}{t}$, L- length and w= width of channel), which shows an exponential drop as a function of distance. We start the experiment at 300 K and cool the specimen at 0.3 K/min to 5 K. The data is acquired every 30 sec. Then, we start heating the specimen to 150 K at 0.3 K/min.

We observe that the non-local resistances while heating do not follow the cooling curve. At 150 K, we raised the temperature to 200K and start cooling again to 5 K. The non-local resistances now follow the heating curve and do not join the first cooling curve. After cooling to 5 K, we raised the temperature to 300 K. We observe that the non-local resistances merge back to the first cooling curve at approximately 250 K. The thermal hysteresis may originate from the coupling of lattice, magnetic and electronic entropies, magnetic ordering, spin fluctuations and microstructural changes[17]. We propose that the observed thermal hysteretic behavior is due to spin accumulation from pure spin current in p-Si. To verify thermal hysteresis, we carried out a similar temperature-dependent longitudinal resistance measurement as shown in Figure 1 d. We observe a thermal hysteresis in longitudinal resistance as well. However, the thermal hysteresis in longitudinal resistance may be due to temperature lag since the resistance measured during heating is lower than the resistance measured during cooling. We propose that the observed thermal hysteresis in non-local resistance is attributed to spin polarization. But the temperature dependent non-local resistance behavior is in agreement with the longitudinal resistance and no change due to ISHE is observed, which is attributed to the small spin-orbit coupling in Si.

The electrical measurement of spin-dependent behavior requires an efficient spin to charge conversion, which is absent in pure p-Si. We hypothesized that Rashba spin orbit coupling due to structure inversion asymmetry (SIA) may allow efficient spin to charge conversion [18] [19], which has been reported for both p-Si and n-Si[20-23]. To test this hypothesis, we deposited a layer of 1 nm of MgO on the p-Si thin film to have SIA and Rashba spin orbit coupling; the MgO/Si interface is observed to have localized electronic states[24]. We repeated the non-local measurement on a MgO/p-Si specimen. First, we applied current across J1 and measured the non-local resistance across J2, J3 and J4 as a function of temperature as shown in Figure 2 a. We observe that the $R_{NL}$ for J2, J3 and J4 increases rapidly as the temperature is reduced from 300 K to 5 K at 0.4 K/min. We observe increase in non-local resistance $R_{J2}$ from 300 mΩ to 870 mΩ, $R_{J3}$ from 30 mΩ to 120 mΩ and for $R_{J4}$ from 0.03 mΩ to 1 Ω. A diffusive spin current will have the largest values closest to source and will decrease exponentially as a function of distance. Now, J2 is closest and J4 is farthest from the location of applied current. The highest non-local resistance is observed at J4 while the smallest at J3. In addition, the $R_{J4}$ changes the sign twice going from positive to negative at ~292 K and turning positive again at ~90 K. We, then, calculated the non-local resistances using van der Pauw's theorem for J2, J3 and J4 as shown in Figure 2 a. The calculated non-local resistances for MgO/p-Si specimen is in agreement with the measured non-local resistances only at 300 K for J3 (red) and J4 (blue). There is disagreement for in the measured and calculated non-local resistance for J2 (black) even at room temperature as shown in Figure 2 a, which suggest there is an additional contribution. The calculated non-local resistances start to deviate as the temperature is lowered as shown in Figure 2 a. With reduction in temperature, non-local resistance increases as opposed to the longitudinal resistance, suggesting an additional temperature dependent contribution, which may be attributed to either the spin or the charge transport.

(Figure 2)

We then measured the non-local resistances for current applied across J2, J3 and J4 junctions as shown in Figure 2 b-d. When the current is applied across J2, we observe that direction of current changes for J1 and J4. Assuming a diffusive spin Hall effect, the non-local resistance should have opposite signs for $R_{J2}$ for $I_{J1}$ as compared with $R_{J1}$ for $I_{J2}$, which is supported by the measurement. In addition, the sign of $R_{J3}$ should not change, which is confirmed as shown in Figure 2 a-b. However, a sign reversal for $R_{J4}$ is observed when the current is applied across J2 as opposed to when current is applied across J1. The sign reversal is not observed for non-local resistances $R_{J4}$ and $R_{J3}$ for current across J3 and J4 respectively as shown in Figure 2 c-d. In all the measurements, a consistent increase in non-local resistances is observed at low temperatures. Further, assuming spin diffusion due to SHE in p-Si, we can calculate the non-local resistance using following equation proposed by Abanin et al.[14]:

$$R_{NL}(x) = \frac{1}{2}\left(\frac{\beta_s}{\sigma}\right)^2 \frac{w}{\sigma l_s} e^{-|x|/l_s}$$

where $\beta_s$ is spin Hall conductivity, $\sigma$ is electrical conductivity, $w$ is width, $l_s$ is spin diffusion length and $x$ is distance from the source. We observe that we cannot fit the parameters since the non-local resistance cannot be higher at longer distances according to this model. We also measured a linear relationship between the non-local voltages as a function of current on a second device to ascertain the repeatability as shown in Figure 3 a-b, which eliminates any effects due to in-plane heat flow, thermal expansion and resultant local stresses.

(Figure 3)

Mihajlovic et al.[16] reported a non-local resistance measurement (similar to this study) on Au thin films and observed negative non-local resistance, which they attributed to quasi-ballistic charge transport. For the electric current across the J1, we do not observe a negative non-local resistance and for J2, J3 and J4, we observe negative non-local resistance for only one location each as shown in Figure 2 b-d. According to the quasi ballistic transport model, only negative non-local resistance should be observed, which we do not. The observed behavior negates the quasi-ballistic charge transport being the underlying mechanism in this study. In addition, the quasi-ballistic charge transport assumes an exponential decay of as a function of length, which we do not observe. We report a sign change in the non-local resistance when the junctions, across whom the current is applied and measured, are switched, which can only arise from the spin transport and not from quasi-ballistic charge transport. We have also demonstrated that Si thin film without MgO layer on top does not exhibit the enhanced non-local resistance behavior as shown in Figure 1 c. We repeated the measurement on Second Si device without MgO layer and it does not show any enhanced non-local resistance as shown in Supplementary Figure S1. This eliminates inhomogeneous doping concentration induced parasitic charge current. We can confirm that the enhanced non-local resistance does not originate from the charge transport across the p-Si layer. The non-local resistance in p-Si only specimen is negligible at 70 μm away (J4), which mean that the length scale for the ballistic transport has to be larger than 70 μm in MgO/Si device, which is larger than the ballistic transport in graphene[25]. Assuming quasi-ballistic charge transport, we hypothesized that the origin of the observed behavior may lie in MgO/p-Si interface. To test this hypothesis, we made a cut using focused ion beam (FIB) across the width of the channel between J2 and J3 to impede the quasi-ballistic transport at the interface for second device. We then measured the non-local resistance on the FIB cut device as shown in Figure 3 c. The observed behavior is similar as presented earlier in Figure 2. This measurement clearly demonstrates that the interfacial charge/spin transport is not the underlying cause of the observed enhanced non-local resistance behavior. The enhanced non-local resistance behavior is repeated for the temperature dependent measurement on third device (Supplementary Figure S2). In order to uncover the effect of doping, we fabricated devices from a SOI wafer with devices layer resistivity of 0.01-0.05 Ω cm (a factor of 10 higher as compared to the first set of devices) and 1 nm MgO top layer. The non-local resistance measurement shows a giant increase in non-local resistance as a function of temperature as shown in Figure 3 d similar to earlier results. These measurements suggest that non-local behavior persists for lower doping levels as well.

We, then, measured the temperature-dependent longitudinal and transverse resistance of MgO/p-Si specimen to parse the spin and charge behavior in non-local measurements as shown in Figure 4 a. From the longitudinal resistance measurement, we observe a metallic behavior in Si and MgO layer at the surface has no effect. However, the measured transverse resistance shows an increase in resistance below ~30 K. The transverse resistance will have contribution from longitudinal resistance (due to misalignment of the Hall bar) and contribution from the other phenomena (probably spin transport). We subtracted the contribution of longitudinal resistance to extract the probable spin transport behavior in transverse resistance as shown in Figure 3 a (inset). We observe a behavior similar to the non-local resistance. This increase in transverse resistance can be attributed to anomalous Hall effect (AHE) or spin accumulation/polarization. To discover the mechanism, we carried out the magnetic field-dependent transverse resistance measurement at 300 K, 200 K, 30 K, 20 K and 5 K Figure 4 b. These measurements show an ordinary Hall effect behavior and we do not observe any anomalous Hall effect. We can deduce that the observed increase in transverse resistance at low temperature is attributed to the spin polarization. In addition, the transverse resistance shows a sign change at 5 K as compared to higher temperatures. This sign change is attributed to the spin accumulation and ISHE at the MgO/p-Si interface towards transverse resistance is greater than the opposing contribution of longitudinal resistance due to misalignment of Hall bar. In a recent study, thermal spin galvanic effect due to Rashba effect has been reported in $Ni_{80}Fe_{20}$/p-Si bilayer thin films[26]. The MgO layer may lead to structure inversion asymmetry resulting in inverse spin-galvanic effect, which may be the underlying cause of transverse resistance behavior. It needs to be pointed out that SHE should not lead to transverse resistance.

(Figure 4)

The intrinsic SHE has been proposed to exist only for p-Si. This behavior necessitates existence of spin-orbit coupled band structure as proposed by Murakami et al.[10]. In Si, only valence band is spin orbit coupled; the conduction band is not. We decided to undertake a temperature-dependent non-local measurement on the MgO (1nm)/ n-Si (2 μm) thin film specimen. The specimen is cooled from 300 K to 5 K and heated to 100 K at 0.4 K/min during acquisition to confirm the reproducibility. Surprisingly, we observe an increase in transverse and non-local resistance behavior similar to that of the MgO/p-Si specimen as shown in Figure 4 c-d. The temperature-dependent longitudinal and transverse resistance is shown in Figure 4 c. The transverse resistance at room temperature is measured to be ~7 Ω, which is very large and cannot be explained by the misaligned Hall bar structure. The non-local resistances are measured by applying current across the J1 as shown in Figure 4 d. In the case of MgO/n-Si, the spin mediated non-local resistance is an order of magnitude larger than the MgO/p-Si. We propose that the observed behavior arises due to spin polarization in n-Si. Since, the transverse resistance is high in the first measurement, we repeated the measurement on second device (Supplementary Figure S3). The transverse resistance is measured to be ~2 Ω at room temperature, which is high as well, and temperature dependent behavior is similar to the first device. We then carried out the transverse resistance measurement on n-Si device without MgO layer as shown in Supplementary Figure S4. We do not observe enhanced transverse resistance and the temperature dependent transverse resistance behavior is similar to longitudinal resistance. This corroborates the enhanced transverse resistance originates from the MgO/n-Si interface similar to MgO/p-Si case. The transverse resistance increases at lower temperatures in case of MgO/p-Si as compared to MgO/n-Si specimen. Based on the experimental observations, we propose that the increase in non-local resistance in case of MgO/Si specimens is attributed to the spin polarization in Si.

Since the spin-orbit coupling in Si is negligible, we hypothesize that spin to charge conversion occurs due to Rashba SOC at MgO/Si and not intrinsically in Si.

(Figure 5)

To further support our argument of the spin current, we recorded the non-local resistance at 5 K as a function of the applied magnetic field in the y and z-direction as shown in Figure 5 a-d. The magnetic field is swept from 8T/-8T while the current is applied across J1. We observe a behavior similar to Hanle precession for $R_{J3}$ for both y-direction and z-direction magnetic field although for large applied magnetic fields as shown in Figure 5 b, d. However, we do not observe Hanle precession in case of $R_{J2}$ as shown in Figure 5 a, c even though J2 is closer to spin source as compared to J3. We repeated the measurement on another device and a similar behavior is observed. From the non-local resistance, Hall resistance and non-local magnetoresistance measurements, we deduce that the observed behavior can be attributed to spin current/polarization but it does not arise from conventional diffusive spin transport. In addition, the observed spin behavior does not arise in the absence of MgO layer.

To understand the role of MgO, we characterized the MgO/Si interface using x-ray photoemission spectroscopy (XPS) as shown in Figure 6 a-c and analyzed using NIST XPS database[27]. X-ray photoelectron spectroscopy (XPS) characterization was carried out by using a Kratos AXIS ULTRA$^{DLD}$ XPS system equipped with an Al Kα monochromated X-ray source and a 165-mm mean radius electron energy hemispherical analyzer. Vacuum pressure was kept below 3 Å~ $10^{-9}$ torr during the acquisition, and the data is acquired at a step of 0.1 eV and dwell of 200 ms. We carried out 10 min of Ar milling to remove the native oxide before sputtering MgO. However, we observe the $Si_{2p}$ peak corresponding to silicon oxide. The analysis of $Mg_{2p}$ reveals a peak corresponding to Mg (51.1 eV) in MgO/Mg as shown in Figure 6 b[28]. From the $Mg_{1s}$ XPS data, we observe a peak corresponding to Mg in $MgSi_2O_4$ (1304.2 eV) and MgO (1303.9 eV). While the XPS data seems inconclusive, we propose that the MgO thin film may have excess oxygen. The excess oxygen in amorphous MgO give rise to ferromagnetic behavior[29-32]. The ferromagnetic MgO layer on the Si may give rise to proximity induced spin splitting and structural inversion asymmetry(SIA). The SIA at the MgO/Si interface will lead to Rashba spin-orbit coupling (SOC). The Rashba SOC mediated spin-Hall magnetoresistance has been reported in nanoscale p-Si (400 nm)[22] and n-Si(2 μm)[23]. The Rashba SOC due to ferromagnetic proximity has been reported for spin-Seebeck effect measurement in $Ni_{80}Fe_{20}$/p-Si bilayers as well[26]. The observed behavior is scientifically significant since intrinsic spin-orbit coupling in Si, O and Mg is small individually, but a combined effect is significant for spin to charge conversion.

(Figure 6)

All the measurement data presented in this study provides an indirect proof of spin current. The SHE leads to spin-orbit torques (SOT) that has been used for magnetization switching[33-39]. In order to uncover the SHE in Si, we decided to undertake experimental measurement of the effect of SOT using anomalous Hall effect (AHE) as a function of applied current. We choose a Co/Pd multilayer specimen with perpendicular magnetic anisotropy (PMA) for this study. We deposited {Co(0.35 nm)/Pd (0.55 nm)}$_5$/ MgO(1nm) on the p-Si Hall bar device using sputtering. We approximate the electrical resistance of the Co/Pd multilayer to be 9 times the p-Si layer. We measure the AHE for an applied electrical current of 1 mA, 5 mA, 10 mA, 12.5 mA and 10 μA as shown in Figure 4 d. We observe a coercive field of 150 Oe at 2 mA of applied current. The coercive field reduces to 100 Oe at 12.5 mA, which changes to 150 Oe as the current is reduced to 10 µA. We clearly observe a reduction in the coercive field of the Co/Pd multilayer thin film. The SHE mediated SOT leads to reduction in the coercive field of the ferromagnetic thin film followed by magnetization switching[39-41]. This work does not explore the quantitative measurement of the SOT but the reduction in coercive field provides a qualitative proof of SHE in Si. It needs to be pointed out that this coercivity reduction is not due to heating since heating reduces magnetic moment as well, which we do not observe. In addition, heating induces permanant microstructural changes in multilayer Co/Pd thin films whereas we observe a complete recovery when the current is reduced.

The Rashba spin-orbit coupled 2DES systems are proposed to exhibit SHE,[42] which has been experimentally observed[43]. In our experimental setup, spin current may originate from the interfacial 2DES. But such systems will have very short spin diffusion length[43] whereas we observe spin transport behavior at a distance of 100 µm. The experiment on device having a FIB cut across the width of the channel directly refutes the interfacial spin or charge transport. Further, the spin current from the MgO/Si interface will not lead to the transverse resistance behavior presented in Figure 4 a and 4 c. The Si and MgO interface has been demonstrated to have interfacial electronic states. The SIA at the MgO/p-Si interface leads to Rashba spin-orbit coupling and efficient spin to charge conversion reported in this work. Lesne et al. experimentally demonstrated the highly efficient spin to charge conversion at oxide interfaces[18]. In addition, $IrO_2$ has been observed to show large spin Hall conductivity [19]. However, the interfacial spin to charge conversion presented is this work involves atoms having insignificant spin orbit coupling individually. We provide a conclusive proof that the spin current originates from the Si and spin to charge conversion takes place at the MgO/Si interface due to Rashba spin orbit coupling.

These measurements led us to believe that the mechanism proposed by Murakami et al.[10] is not the underlying mechanism for the observed behavior since SHE has not been predicted for n-Si. Zhang et al.[44] theoretically predicted that the site inversion asymmetry in Si may create hidden spin polarization. The lattice of Si is centosymmetric, but individual sites are not. This leads to intrinsic spin polarization, which is hidden due to the compensation by the inversion counterpart. This behavior can be regarded as local antiferromagnetic like non-equilibrium spin polarization[45]. This behavior has been supported by recent observation of spin mediated emergent antiferromagnetic phase transition in both n-Si and p-Si[20, 22, 23]. We propose that the SHE is attributed to the site inversion asymmetry of diamond cubic lattice. The MgO layer creates Rashba SOC in the Si layer in addition to intrinsic site inversion asymmetry, which leads to helical spin states in Si, generating dissipationless spin current observed in this study. The Rashba SOC at the MgO/Si interface leads to efficient spin to charge conversion. The Rashba SOC at the MgO/Si interface may also lead to spin-galvanic effect and in turn spin polarization in the Si layer, which leads to the anomalous increase in Hall voltage at the low temperatures. The SHE and helical spin transport behavior is supported by the change in sign of non-local resistances presented in the Figure 2-3.

In conclusion, we report generation of dissipationless spin current in a Si specimen without any ferromagnetic source and efficient spin to charge conversion by having structure inversion asymmetry at the MgO/Si interface. The dissipationless helical spin current originates from the site inversion asymmetry in a centosymmetric diamond cubic lattice of Si. The spin current leads to spin-orbit torques and reduction in coercivity of Co/Pd multilayer specimen. These results will lay the foundation of semiconductor spintronics without ferromagnetic spin source and will also advance the spin transport characterization techniques. In addition, these results will lead to advancement of interfacial spin to charge conversion.

**Acknowledgement**

We thank Prof. Ward Beyermann (UCR) for discussions and inputs. The XPS facilities used in this work is supported by NSF grant DMR-0958796.

[38]  Lau Y-C, Betto D, Rode K, Coey J M D and Stamenov P *Nat Nano* **11** 758-62 (2016)

[39]  Fukami S, Zhang C, DuttaGupta S, Kurenkov A and Ohno H *Nat Mater* **15** 535-41 (2016)

[40]  Garcia D, Lou P C, Butler J and Kumar S *Solid State Communications* **246** 1-4 (2016)

[41]  Akyol M, Yu G, Alzate J G, Upadhyaya P, Li X, Wong K L, Ekicibil A, Khalili Amiri P and Wang K L *Applied Physics Letters* **106** 162409 (2015)

[42]  Bernevig B A and Zhang S-C *Phys. Rev. Lett.* **95** 016801 (2005)

[43]  Choi W Y, Kim H-j, Chang J, Han S H, Koo H C and Johnson M *Nat Nano* **10** 666-70 (2015)

[44]  Zhang X, Liu Q, Luo J-W, Freeman A J and Zunger A *Nat Phys* **10** 387-93 (2014)

[45]  Jungwirth T, Marti X, Wadley P and Wunderlich J *Nat Nano* **11** 231-41 (2016)
18

List of Figures:

Figure 1 a. Hypothesis of the experimental work, b. the experimental setup with the freestanding p-Si specimen, c. temperature-dependent non-local resistance across J4 and J2 while the electrical current is applied across J3, and d. the temperature-dependent longitudinal resistance of the p-Si specimen.

Figure 2. The non-local resistances as a function of temperature for an applied current across a. J1, b. J2, c. J3 and d. J4.

Figure 3. The non-local resistance as a function of current applied across J3 and non-local resistance measured across a. J2, b. J4, c. the non-local resistance as a function of temperature for the specimen with FIB cut between J2 and J3, and d. the non-local resistance measurement as a function of temperature for lightly doped p-Si specimen.

Figure 4 a. Temperature-dependent longitudinal and transverse resistance of MgO/p-Si specimen, transverse resistance as a function of temperature after subtraction of longitudinal contribution (inset), b. the transverse resistance as a function of magnetic field at 300 K, 200 K, 30 K, 20 K and 5 K, c. temperature-dependent longitudinal and transverse resistance of MgO/n-Si specimen and d. the non-local resistance as a function of temperature for an applied current across J1 of MgO/n-Si specimen.

Figure 5. The non-local resistance measurement as a function of magnetic field for the current applied across J1 and non-local resistance measured across J2 a. field along y axis, c. field along z-axis and non-local resistance measurement across J3 b. field along y-axis and d. field along z-axis.

Figure 6. a. the XPS spectrum corresponding to the $Si_{2p}$, b. the XPS spectrum corresponding to the $Mg_{2p}$, c. the XPS spectrum corresponding to the $Mg_{1s}$ and d. the anomalous Hall resistance measurement as a function of electric current showing reduction in coercivity of Co/Pd multilayer due to spin-orbit torques.

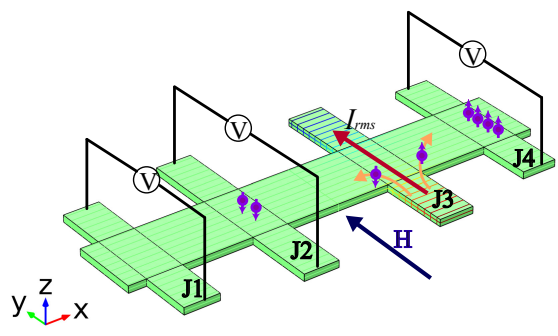
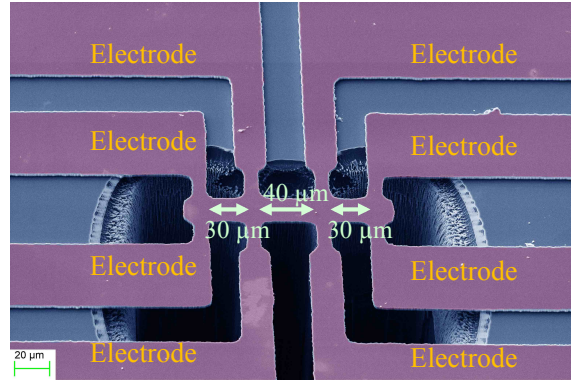
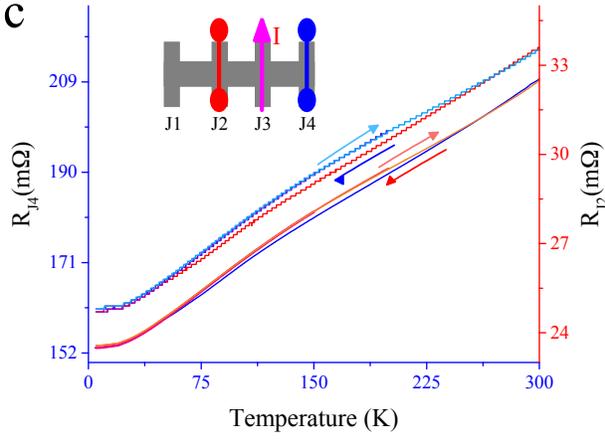
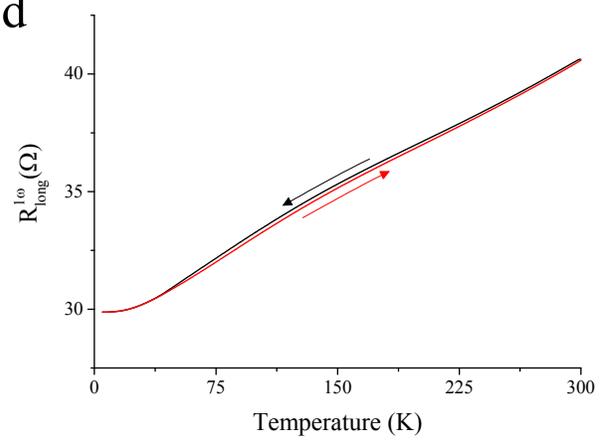

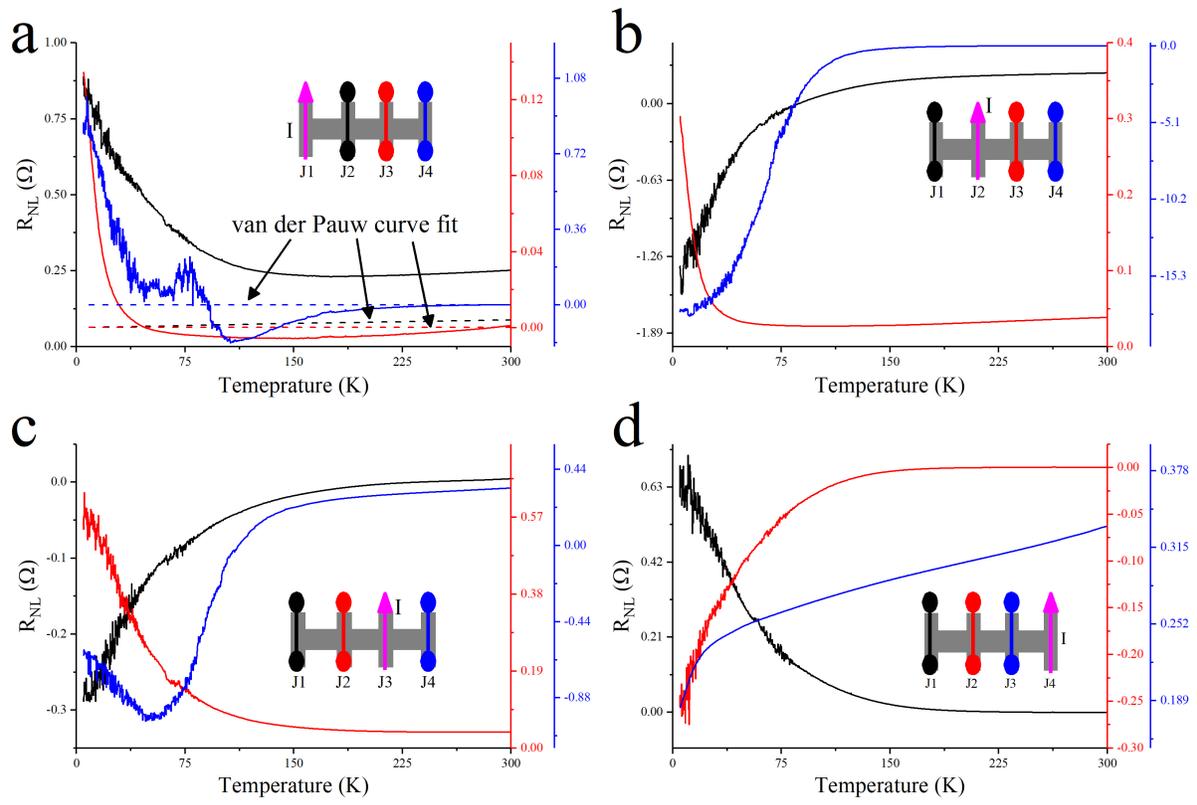

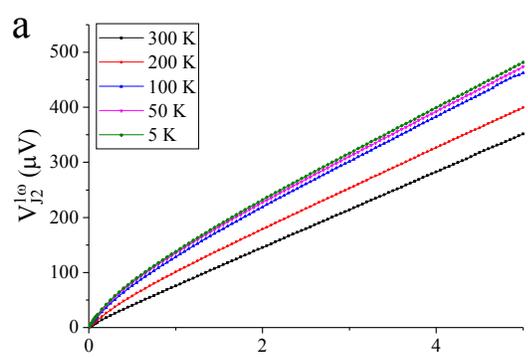
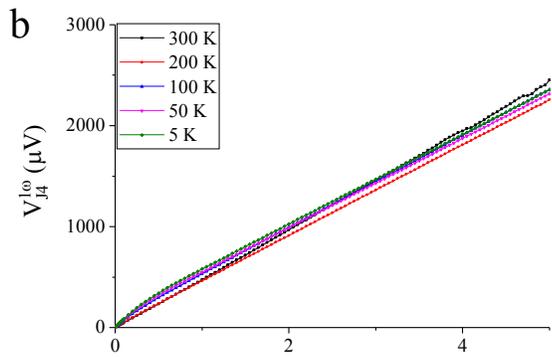
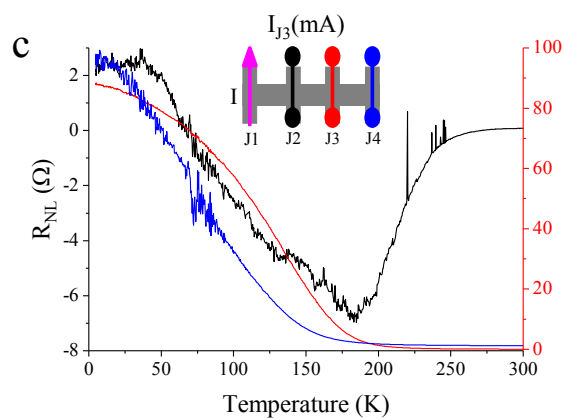
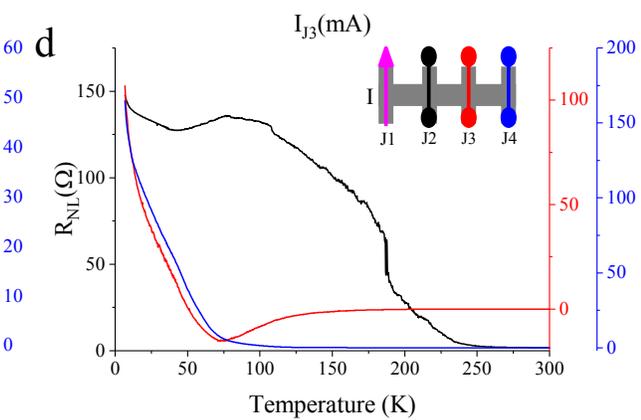

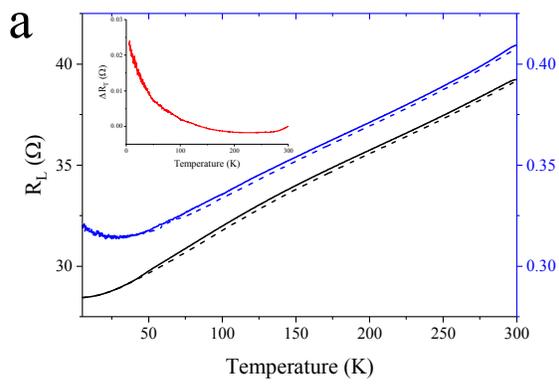
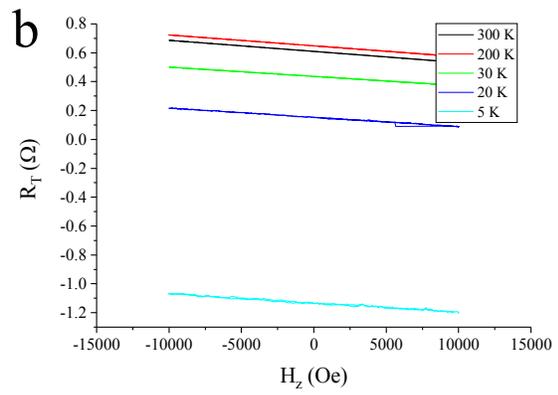
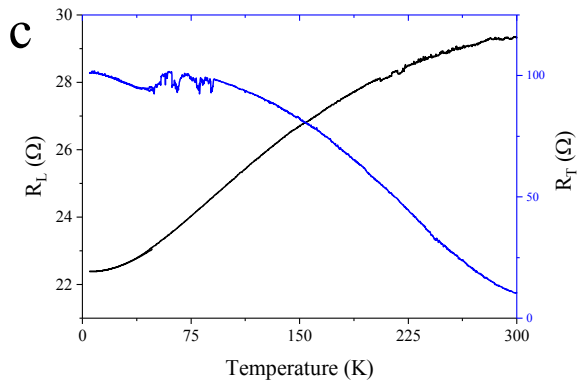
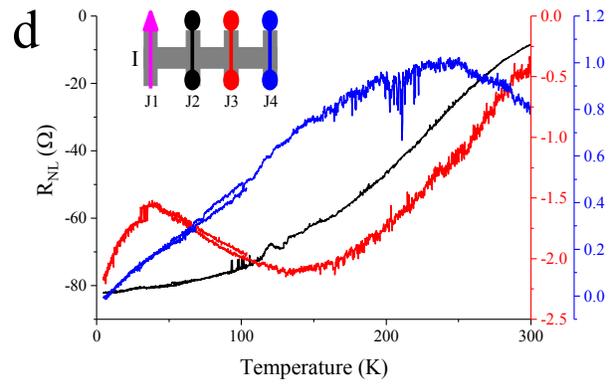

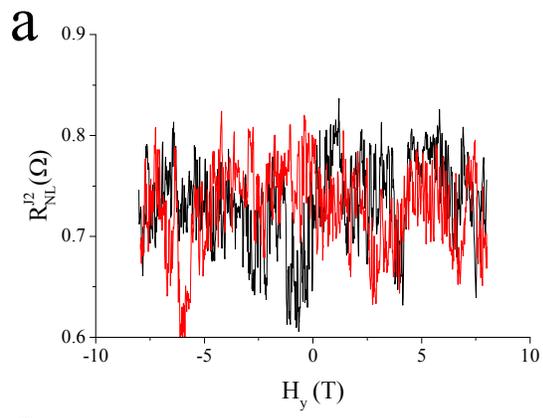
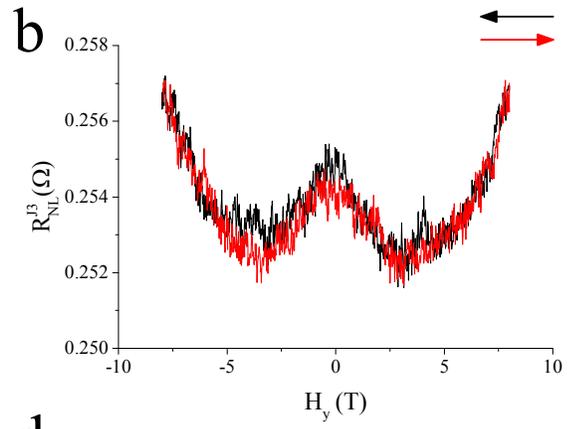
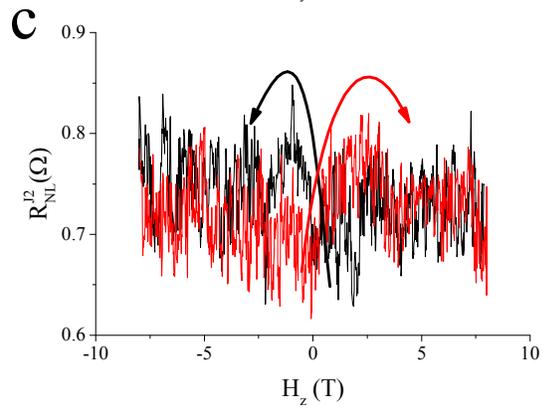
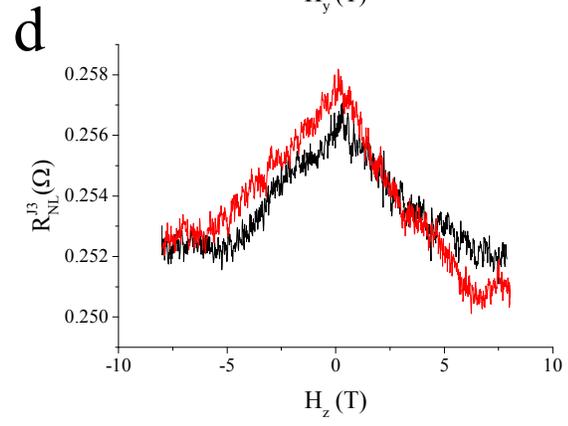

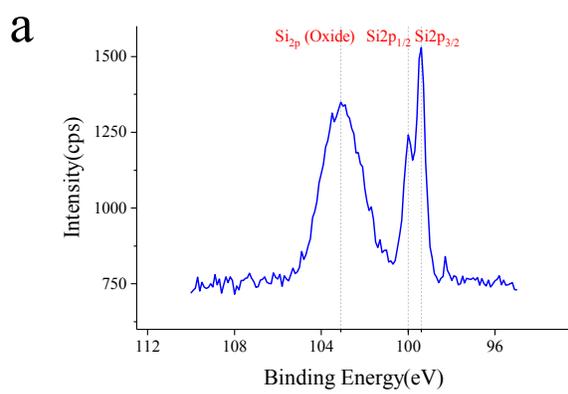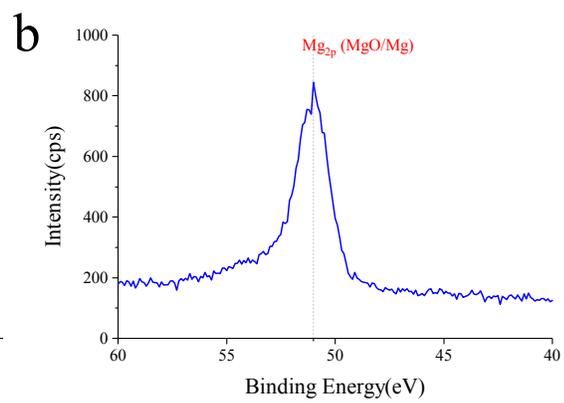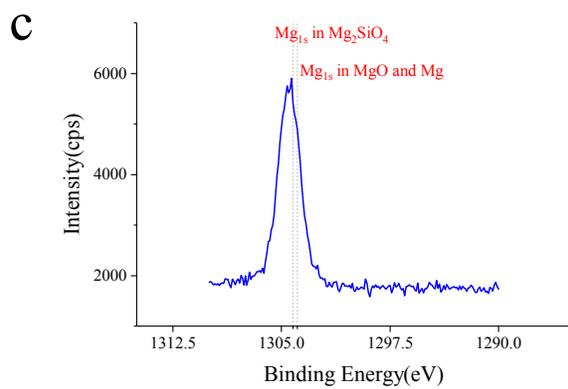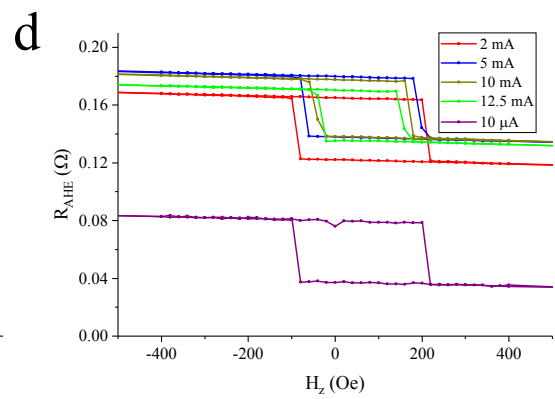

**Supplementary Material-**

Supplementary Figures

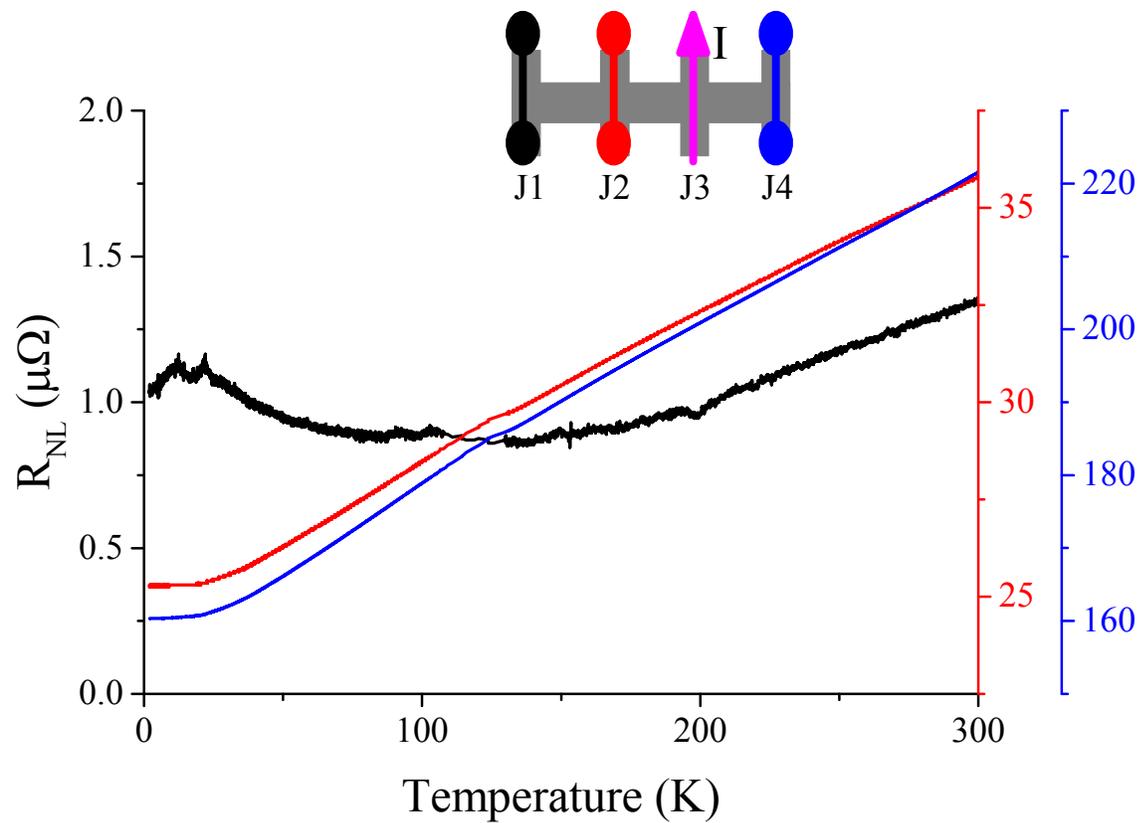

Supplementary Figure S1. The non-local resistance of Si only device (Without MgO) as a function of temperature measured across J1, J2 and J4 while the electrical current is applied across J3.

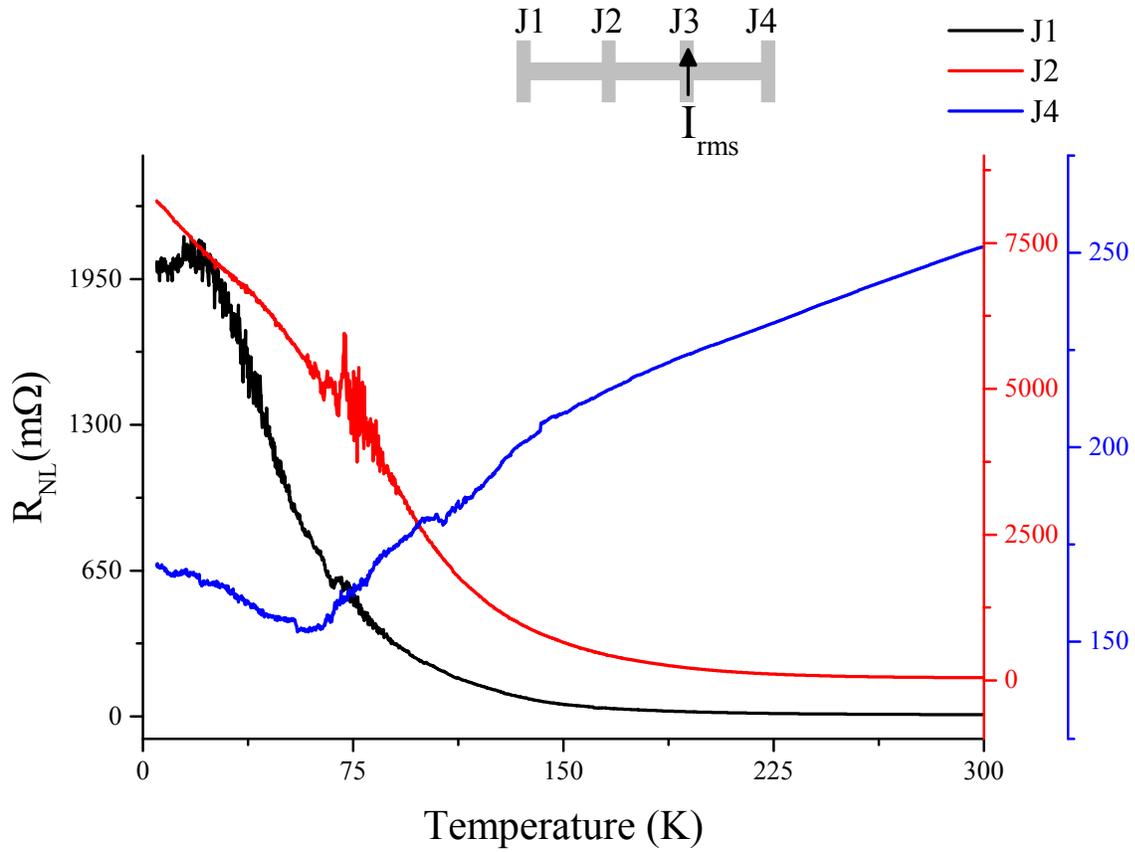

Supplementary Figure S2. The non-local resistance of MgO/p-Si third specimen for current applied across J3 and resistance measured across J1, J2 and J4.

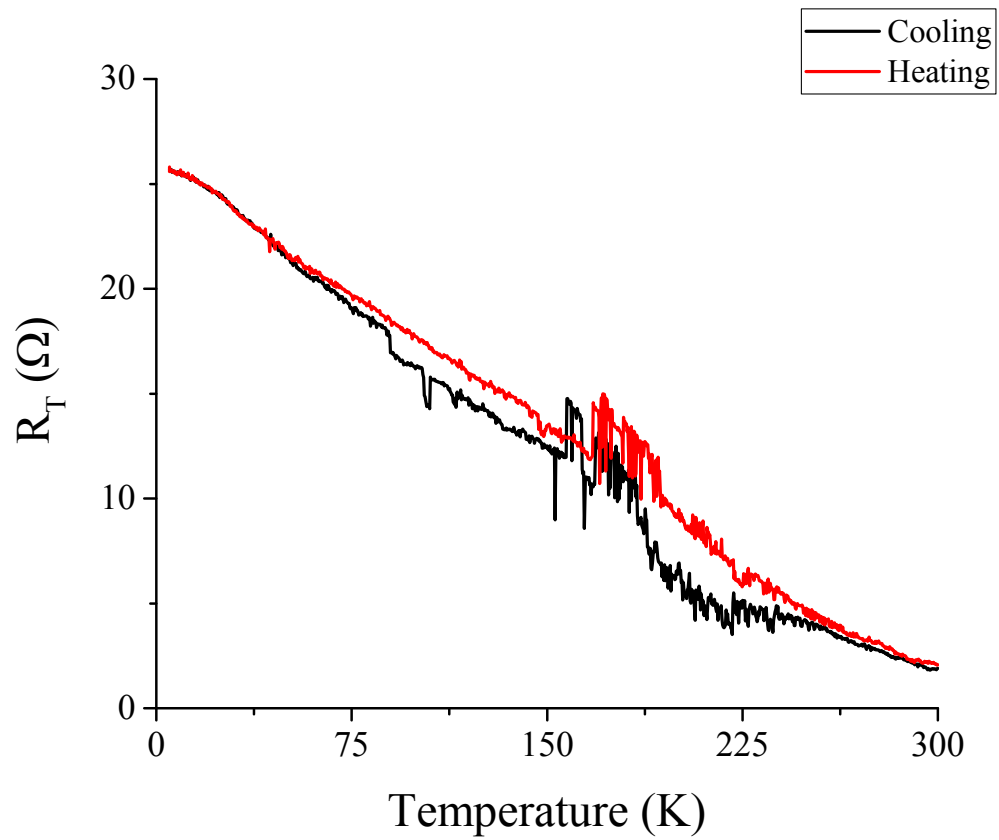

Figure S3. The temperature dependent transverse resistance of MgO/n-Si specimen.

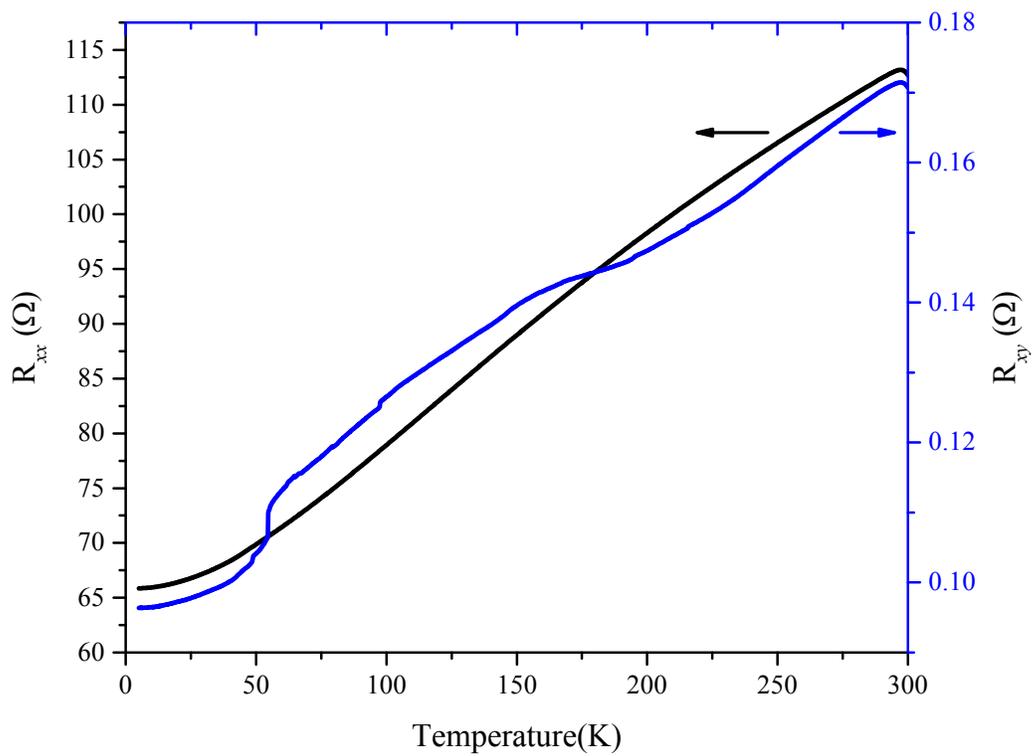

Supplementary Figure S4. Temperature-dependent longitudinal and transverse resistance of n-Si specimen without MgO.